\documentclass[fleqn,10pt]{wlscirep}
\title{Theory of Kerr and Faraday rotations and linear dichroism in Topological Weyl Semimetals}

\def\be{\begin{equation}}
\def\ee{\end{equation}}

\def\bi{\begin{itemize}}
\def\ei{\end{itemize}}
\def\bn{\begin{enumerate}}
\def\en{\end{enumerate}}
\def\bea{\begin{eqnarray}}
\def\eea{\end{eqnarray}}

\def\ba{\begin{array}}
\def\ea{\end{array}}
\def\bd{\begin{displaymath}}
\def\ed{\end{displaymath}}

\author{Mehdi Kargarian}
\author{Mohit Randeria}
\author{Nandini Trivedi}
\affil{Department of Physics, The Ohio State University, Columbus, OH 43212, USA}

\begin{abstract}
We consider the electromagnetic response of a topological Weyl semimetal (TWS) with 
a pair of Weyl nodes in the bulk and corresponding Fermi arcs in the surface Brillouin zone. 
We compute the frequency-dependent complex conductivities $\sigma_{\alpha\beta}(\omega)$
and also take into account the modification of Maxwell equations by the topological $\theta$-term
to obtain the Kerr and Faraday rotations in a variety of geometries.
For TWS films thinner than the wavelength, the Kerr and Faraday rotations, determined by the separation between Weyl nodes,
are significantly larger than in topological insulators.
In thicker films, the Kerr and Faraday angles can be enhanced by 
choice of film thickness and substrate refractive index.
We show that, for radiation incident on a surface with Fermi arcs, there is no Kerr or Faraday rotation
but the electric field develops a longitudinal component inside the TWS, and there is linear dichroism signal.
Our results have implications for probing the TWS phase in various experimental systems.
\end{abstract}
\begin{document}
\flushbottom
\maketitle
\thispagestyle{empty}


\section*{Introduction}
In recent years, condensed matter physics has witnessed the emergence of novel quantum phases characterized by topology rather than by symmetry breaking. 
The best studied of these are the topological insulators, which have a bulk band gap, but with gapless edge or surface states protected by time reversal symmetry
and characterized by topological invariants~\cite{Hasan:rmp10,Qi:rmp11,HasanMoore:Ann11}. 
More recent predictions suggest that nontrivial topological properties can also arise in
certain systems whose gapless band structures are characterized by point or line nodes~\cite{volovik:book,vishwanath:prb11,burkov:prb11,vafek:annal14}. A particularly interesting state of matter is the topological Weyl semimetal (TWS)~\cite{vishwanath:prb11,burkov:prb11,yingran:prb11}.
These are phases with broken time-reversal or inversion symmetry, whose electronic structure consists of pairs of Weyl nodes, points in the bulk Brillouin zone (BZ), which are at the chemical potential and act as sources and sinks of Berry curvature. This is predicted~\cite{vishwanath:prb11} to lead to 
unusual surface states that are gapless on disconnected Fermi arcs with end points at the projections of the bulk nodes onto the surface BZ.

A number of candidate material systems that should exhibit a TWS phase have been proposed. These include 
pyrochlore iridates A$_2$Ir$_2$O$_7$~\cite{vishwanath:prb11}, spinels~\cite{xu:prl11} and 
multilayers of topological insulators and trivial insulators~\cite{burkov:prl11}, all of which break time reversal. There is a recent report of the 
experimental observation~\cite{xu:arxiv1502.03807} of surface Fermi arcs in TaAs, which is a TWS by breaking spatial inversion. 
In addition there are many theoretical predictions about the unusual transport and magnetic properties of 
TWSs~\cite{yingran:prb11,burkov:prl11,hosur:prl12,chen:prb13,Kim:prl13,son:prl12,vazifeh:prl13,chen:prb13,
Parameswaran:prx14,hosur:prb15,Goswami:arxiv:1404.2927}.   

In this paper, we theoretically address the electrodynamic response of a TWS with broken time reversal symmetry,
focusing on Kerr and Faraday rotations and linear dichroism.
We show that our predictions are sensitive to four nontrivial characteristics of a TWS:
(i) the topological Weyl nodes which lead to nontrivial $\sigma_{xy}$ in the absence of an applied field,
(ii) the nodal excitations leading to optical conductivity $\sigma_{xx}(\omega) \sim \omega$,
(iii) the unusual surface states with Fermi arcs, and
(iv) the modification of Maxwell equations inside the TWS via a topological $\theta$-term.

\begin{figure}[ht]
\centering
\includegraphics[width=\linewidth]{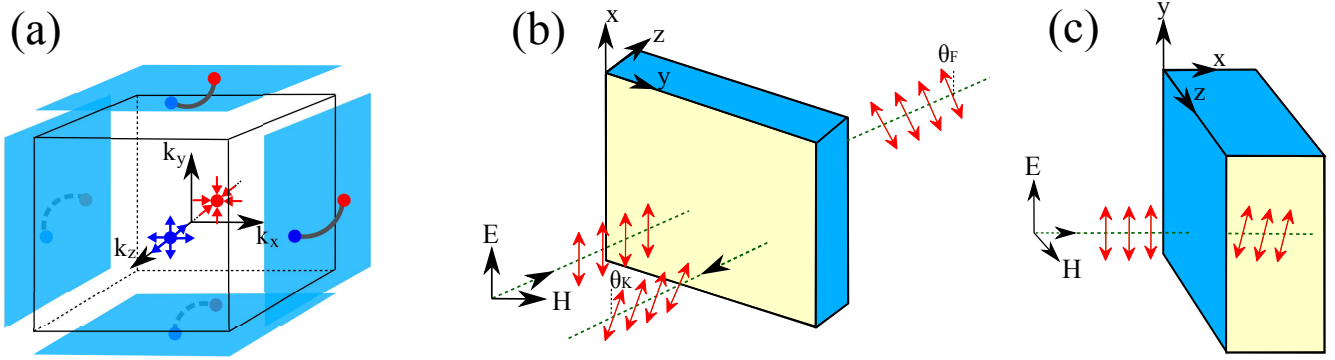}
\caption{(Color online) (a) Schematic {\bf k}-space picture showing the location of Weyl nodes (blue and red dots with outward arrows as sources and inward arrows as sinks of Berry curvature) along $k_z$ axis in bulk Brillouin zone (BZ)
and the Fermi arcs (lines ending at the projection of the Weyl nodes) on the surface BZ's, shown as blue squares.  Note that surfaces perpendicular to $z$ axis have no arcs. 
(b, c) Real-space geometry of slab and electromagnetic waves. The blue (dark) surfaces of slabs support
the arc surface-states and yellow (light) surfaces are ones without surface states. 
In (b) we show the Kerr $\theta_K$ and Faraday $\theta_F$ angles. 
For geometry shown in (c) there are no Kerr and Faraday rotations, instead ${\bf E}$ acquires a 
longitudinal component inside the Weyl semimetal.}
\label{slab_arc_FK}
\end{figure}

In brief, our results are as follows.
We find that there is a very important difference between the EM responses for radiation normally incident on 
(A) a surface that does not support Fermi-arc electronic states, versus (B) a surface that does.
Specifically, for a pair of nodes separated along the $k_z$-direction in the bulk BZ, the $(x,y)$-plane has no Fermi
arc states; see Fig.~\ref{slab_arc_FK}. 
We find Kerr and Faraday rotations only in case (A), i.e., for light incident on the $(x,y)$-plane.
For case (B), there is no Kerr and Faraday rotations, but the electric field develops a longitudinal component inside the TWS
and there is a finite magnetic linear dichroism signal.

We look at various experimental geometries -- a TWS film thinner than the wavelength of light, a semi-infinite slab
and a film with thickness comparable to wavelength -- and identify cases where large
measurable signals are obtained.  
We discuss at the end of the paper, various experimental systems where our results can be tested.
In addition to the TWS materials already mentioned above,
our results may also be generalized to recently discovered Dirac semimetals, where a magnetic field
${\bf B}$ separates the Weyl nodes.

\section*{Results}
\subsection*{The conductivity tensor}
The first step in computing the Faraday and Kerr effects for a TWS is to 
obtain the conductivity tensor $\sigma_{\alpha\beta}(\omega)= \sigma_{\alpha\beta}'+i\sigma_{\alpha\beta}''$,
using the Kubo formalism. To make the physical ideas clear, we focus on a TWS with
two Weyl nodes with opposite chiralities, located at $\pm{\bf b} = \pm(0,0,b)$.
The low-energy physics is described
by the linearized Hamiltonian 
 \bea \label{H_linear} 
 H(\textbf{k})=\pm \hbar v_F\vec{\sigma} \cdot (\bf{k}\pm\bf{b}),
\eea  where $k \leq k_c$, the momentum cut-off, $v_F$ is the Fermi velocity, and $\vec{\sigma}=(\sigma_x,\sigma_y,\sigma_z)$ are Pauli matrices. 

In addition to the known result~\cite{hosur:prl12} $\sigma_{xx}'(\omega) =  e^2|\omega|/6h v_F$,
we also determine $\sigma_{xx}''(\omega)$. 
While the linearized $H$ is sufficient to obtain $\sigma'(\omega)$ for $\omega \ll \omega_c=v_F k_c$, 
it leads to a pathology in $\sigma''(\omega)$, even in the low frequency regime. This is avoided by using a lattice regularization.
We find that $\sigma_{xx}''$ has a spurious $-1/\omega$ term for the linearized dispersion, which
is in fact exactly cancelled by the diamagnetic term in the Kubo formula.
Such a diamagnetic term is indeed present for the lattice $H$, but absent for a strictly linear dispersion; see Methods for more details.

Using $\epsilon=\tilde{\epsilon}/\epsilon_0=\epsilon_b+i\sigma_{xx}/\epsilon_0\omega$,
where $\epsilon_b$ is the bound charge contribution,
we then find the real and imaginary parts of the dielectric function.
These are given by
$\epsilon'(\omega)=\epsilon_{b}+ (\alpha c/3\pi v_F)\ln\left|{(\omega^2-4\omega_c^2})/{\omega^2}\right|$ and $\epsilon''=\alpha c/3\pi v_F$, 
where $\alpha={e^2}/{4\pi\epsilon_0 \hbar c}$ is the fine structure constant.

For the Hall response, we find the $\omega^2$ correction to the known d.c.~result~\cite{yingran:prb11,burkov:prl11}
and obtain
$\sigma_{xy}'(\omega) = {e^2 b}/{\pi h}+ (e^2b/6\pi h)\omega^2 / (\omega_c^2 - v_F^2 b^2)$.
The $\omega^2$ terms in $\epsilon'$ and $\sigma_{xy}'$ are shown here
for the linear model; for the lattice model the prefactors, which must be found numerically, 
have essentially the same structure with $\omega_c \sim v_F/a$.
We note that disorder produces subleading corrections~\cite{hosur:prl12} to $\sigma_{xx}'$,
and does not~\cite{Burkov:prl14} affect $\sigma_{xy}'$ . 
Therefore the above results remain valid at least for weak disorder. 

\subsection*{Electromagnetic Response}
We now consider in turn the electromagnetic response of a TWS with light incident on:
(A) a surface that does not support Fermi-arc states, which is the $(x,y)$-plane
with nodes separated along the $z$-direction, and
(B) surfaces that do support Fermi-arc states; see Fig.~\ref{slab_arc_FK}.
We look at three geometries: (1) In an ultra-thin film, which is thinner than the wavelength, 
the $\sigma_{\alpha\beta}(\omega)$ of TWS impacts the boundary condition
at the interface between two topologically trivial media.
(2) When the TWS is a semi-infinite slab, we need to consider the modification in
the Maxwell equations arising from the topological $\theta$-term.
(3) In a film of finite thickness comparable to the wavelength of light, we take into
account both the modified Maxwell equations and interference phenomena arising
from multiple reflections. 

\subsubsection*{Kerr and Faraday rotations}
\begin{itemize}
\item Case (A1): We first consider light incident on the $(x,y)$ surface with 
no arc states [see Fig. \ref{slab_arc_FK}(b)] for an ultra-thin film with thickness $d$ with
$a\ll d\ll \lambda$, the wavelength of light. In this limit, the nontrivial properties of the 
TWS enter only through the boundary condition for e.m.~fields
in the two non-topological media on either side ($z\!<\!0$ and $z\!>\!0$). 
The surface current density at $z\!=\!0$ is 
$\textbf{J}^s_{\alpha} =\sigma^s_{\alpha\beta} \textbf{E}_\beta$, with the surface
conductivity tensor $\sigma^s_{\alpha\beta} = d \sigma_{\alpha\beta}$ for
the TWS thin film. The rotation of the polarization in both reflection and transmission is  
governed by $\sigma'_{xy}$.

Using right and left circularly polarized
transmitted fields $E^{t}_{\pm}=E^{t}_{x}\pm iE^{t}_{y}$, 
the Faraday rotation is given by $\theta_{F}=(\arg{E^{t}_{+}}- \arg{E^{t}_{-}})/2$.
An analogous treatment of the reflected fields yields the Kerr rotation.
We thus find
\bea \label{theta_F_ultra}
\tan{\theta_F} &=& 2\alpha bd/\pi\left[{2+\alpha\omega d/3v_F }\right]  \\ \label{theta_K_ultra}
\tan{\theta_{K}} &=& \frac{-4\alpha bd/\pi}{(\alpha\omega d/3v_F)^2+2\alpha\omega d/3v_F +(2\alpha bd/\pi)^2}.
\eea
Here we ignore the $\omega$-dependence of $\sigma'_{xy}$ and show results for free standing films. 

At the lowest frequencies, we find
$\tan\theta_F \approx  bd\alpha/\pi = - \cot\theta_K $.
Note that $\theta_F$ for a TWS is enhanced by a factor $bd \gg 1$,
relative to the small ($<1^{\circ}$) Faraday rotation for the
surface of a topological insulator with time reversal breaking~\cite{tse1:prl10,tse2:prb11}. 

We find that $\theta_K$ for a TWS can also be very large. For instance for typical film thickness of $d\!=\!50\!-\!100$~nm it reaches values of $70\!-\!80^{\circ}$.  

We also note that there is a finite frequency regime where the results are independent of film thickness $d$. 
For $6bv_F/\pi<\omega<3v_F/\alpha d$, two second-order terms in denominator in eq.~(\ref{theta_K_ultra})
can be ignored. We thus obtain $\tan{\theta_K}=-6v_Fb/\pi\omega$, independent of $d$. 
For a typical values of $d=100$~nm, $v_F=10^6$~m/s, the corresponding photon energies should be 
$1<\hbar\omega<5$~eV, corresponding to the visible range, which is accessible in experiments. 

\item Case (A2): For a semi-infinite TWS ($z \geq 0$) with light incident 
on a surface without arc states [see Fig. \ref{slab_arc_FK}(b)], we must take into account
the modification of the Maxwell equations inside the TWS. The topological axion term~\cite{wilczek:prl87,zyuzin:prb12}
in the action with $\theta({\bf r}) = {\bf b}\cdot{\bf r}$ leads to
\bea \label{modified_maxwell}
\nabla\cdot \textbf{D}&=&\rho+\kappa\textbf{b}\cdot \textbf{B}, \\ \label{modified_maxwell_H}
\nabla\times\textbf{H}&=&\partial \textbf{D}/\partial t+\textbf{J}-\kappa \textbf{b}\times \textbf{E}
\eea
with
$\kappa=(2{\alpha}/{\pi})\sqrt{{\epsilon_0}/{\mu_0}}$=$e^2/\pi h$. 

In our geometry [Fig.~\ref{slab_arc_FK}(b)] the $\textbf{E}$ and $\textbf{B}$ fields are in the $(x,y)$ plane.
As usual, $\textbf{B}\!=\!\tilde{\mu}\textbf{H}$  with $\mu=\tilde{\mu}/{\mu_0}=1+\chi_m$.
The effective dielectric tensor (${\alpha,\beta}=x,y$) of the TWS can be written as
\bea
\label{e_tensor_txt} 
{\epsilon}_{\alpha,\beta}=\epsilon\delta_{\alpha,\beta}-(2{\alpha}/{\pi})({c}b/{\omega})\left(\tau^{y}\right)_{\alpha,\beta}.
\eea 
$\epsilon$ is the complex dielectric function $\epsilon=\epsilon_b+i\sigma_{xx}/\epsilon_0\omega$ and
the Pauli matrix $\tau^{y}$ term arises
from $\textbf{b}\times\textbf{E}$ with $\textbf{b}\!=\!b\hat{\bf z}$.
Thus the constitutive relation for a TWS
is of the gyrotropic form~\cite{LANDAU,Visnovsky} $\textbf{D}=\epsilon\textbf{E}-i\textbf{g}\times\textbf{E}$,
well known in the electrodynamics of ferromagnetic materials, with the nodal separation ${\bf b}$ as
the gyration vector~\cite{LANDAU} $\textbf{g}$.

We use the Jones~\cite{Visnovsky} basis of eigenvectors of $\tau^{y}$ to obtain the complex refractive indices
\bea \label{npm} n^{\pm}=\sqrt{\epsilon \pm \alpha\lambda_0 b/\pi^2},\eea 
for left and right circular polarized light inside the TWS. with $\lambda_0=2\pi c/\omega$
the wavelength in vacuum. The difference between $n^{\pm}$ leads to 
{\it birefrigence}. 

The tangential electric fields are continuous across the vacuum-TWS interface,
because the total surface current density at $z\!=\!0$ vanishes.
First, there is no transverse response localized on the $z\!=\!0$ surface.
Second, in contrast to metals, the longitudinal current density in a TWS is not localized at the surface,
given the long penetration depth $\delta\sim\epsilon'\lambda/\pi\epsilon''\gg\lambda$ over which fields decay 
in the TWS. 

Writing $(n^{-}-n^{+})/(1-n^{+}n^{-}) = \eta e^{i\phi}$, we find the final result for the Kerr rotation to be (see Methods)
\bea \label{KF_semi} \theta_K&=&\frac{1}{2}\tan^{-1}\left(\frac{2\eta\sin\phi}{\eta^2-1}\right)\simeq  \frac{\alpha^{2} c}{6\pi^2 v_F}\frac{\lambda b}{\epsilon'(\epsilon'-1)},
\eea 
where $\lambda\!=\!\lambda_{0}/\sqrt{\epsilon'}$ is the wavelength inside the TWS.

The dielectric constant of TWS can be rather large, e.g., in a topological insulator (TI)-ordinary insulator multilayer
Bi$_2$Se$_3$ has $\epsilon'\!\sim\!30-80$. Therefore the Kerr rotation from a single interface is small $\sim 10^{-3}-10^{-4}$ rad 
(like in TI's~\cite{tse1:prl10,tse2:prb11,karch:prl09,Maciejko:prl10}.) We show next that reflection from a thick film 
can substantially enhance the Kerr rotation.

\item Case (A3): We next consider a TWS slab of thickness $d$ and dielectric function
$\epsilon'_2+i\epsilon_{2}''$ sandwiched between two 
non-topological media with refractive indices $n_1=1$ (vacuum) and $n_3$ (substrate).  
As before, we look at circularly polarized light incident on a TWS surface with no arcs [Fig. \ref{slab_arc_FK}(b)]. 
Using eq.~(\ref{npm}) and $n_2 =\sqrt{\epsilon'_2}$, the refractive index for the TWS is 
given by $n_2^{\pm}\simeq n_2(1+x^{\pm})$, where $x^{\pm}=\pm \alpha \lambda_0 b/2\pi^2 \epsilon'_2+i\epsilon_{2}''/2\epsilon'_2$ 

The multiple scattering from the two interfaces lead to interference effects
with reflection and transmission coefficients that vary with thickness $d$.
Here we mention only some of the results focusing on the cases where a large Kerr/Faraday rotation is predicted. At reflection maxima, where
$d=p\lambda/2$ for an integer $p$, the Kerr rotation for a free standing slab
with $n_{1}= n_{3}$ is given by
\bea\theta_{K}\simeq -\tan^{-1}\left(\frac{\alpha\lambda_{0}b}{\pi^2\epsilon_{2}''}\right),
 \eea 
independent of $\epsilon'$. For THz radiation $\lambda_{0}=10\,\mu$m and Fermi velocity $v_{F}\simeq 10^{6}m/s$, we get large $\theta_{K}\simeq-\pi/2$.
We can also get a large Kerr angle $\simeq\pi/2 - {\pi^2\epsilon_{2}''}/{\alpha\lambda_{0}b}$ near a reflection minima, when $d=(2p+1)\lambda/4$, 
provided one can also satisfy $n_2\simeq\sqrt{n_1n_3}$ by choice of substrate.
The Faraday rotation of transmitted waves can be similarly enhanced for refractive indices 
satisfying $n_{2}\simeq \sqrt{n_1n_3}$, 
with
\bea \theta_{F}\simeq-\frac{\pi}{2}+\frac{\alpha\lambda_{0}b}{8\pi n_2}, \eea
where the second term is very small for wavelength of about $\lambda_0 <10\,\mu$m. 

\end{itemize}
\subsubsection*{Linear dichroism}
Finally we turn to light incident on a surface of the TWS that supports Fermi arcs.
We take the incident wave propagating along $\hat{\bf x}$  [see Fig.~\ref{slab_arc_FK}(c)], 
the Weyl nodes are at $\pm b \hat{\bf z}$ in ${\bf k}$-space with Fermi arcs in the $(k_y,k_z)$ surface Brillouin zone.
We will focus here on Kerr reflection from a semi-infinite TWS ($x\!>\!0$).

Consider the propagation of a plane wave with wavevector $k=n\omega/c$ inside the TWS.
The modified Maxwell equations (\ref{modified_maxwell_H}) lead to
\bea \label{Eyz_txt}
\left( \begin{array}{cccc}
  0\\ 
  n^2E_y\\
  n^2E_z\\  
  \end{array}\right)=\left(
  \begin{array}{cccc}
  \epsilon&i\epsilon_{xy}'&0\\ 
 -i\epsilon_{xy}'&\epsilon&0\\
 0&0&\epsilon\\  
  \end{array}\right)\left( \begin{array}{cccc}
  E_x\\ 
  E_y\\
  E_z\\  
  \end{array}\right),
\eea where $\epsilon=\epsilon_b+i\sigma_{xx}/\epsilon_0\omega$ and 
$\epsilon_{xy}'=2\alpha cb/\pi\omega$t.
In order to have a solution, the refractive index $n$ must be either (i)
$n_{\|}^2=\epsilon$ or (ii) $n_{\bot}^2=\epsilon-\epsilon'^2_{xy}/\epsilon$. 
As we shall see next, the subscript here refers to whether 
the incident ${\bf E}$ field is ${\|}$ or ${\bot}$ to the node separation along $\hat{\bf z}$.
For case (i), $\textbf{E}=\hat{\bf z}E_{0}e^{i\omega(\pm n_{\|}x/c-t)}$
and matching boundary conditions at $x\!=\!0$, we find
no rotation of $\textbf{E}$ upon entering the TWS. This is easy to see
since $\textbf{b}\cdot\textbf{B}$ and $\textbf{b}\times\textbf{E}$ in eq.~(\ref{modified_maxwell}) 
are identically zero, when ${\bf E}||\hat{\bf z}$ and ${\bf H}||\hat{\bf y}$.

Case (ii), relevant for incident ${\bf E}||\hat{\bf y}$ and ${\bf H}||\hat{\bf z}$  [see Fig.~\ref{slab_arc_FK}(c)],
is more interesting. Now the solution inside the TWS is $\textbf{E}=\left(\hat{\bf y}-i \hat{\bf x}\ {\epsilon'_{xy}}/{\epsilon} \right)E_{0}e^{i\omega(\pm n_{\bot}x/c-t)}$. 
Thus we find no Faraday or Kerr rotation of the polarization in the $(y,z)$-plane, however,
we do see that $\textbf{E}$ acquires a {\it longitudinal} component inside the TWS.
These conclusions, which are further elucidated below, can be reached just by looking at the solution of eq.~(\ref{modified_maxwell}) 
inside the TWS ($x\!>\!0$). The boundary conditions at $x\!=\!0$ (discussed in Methods) only determine the magnitude of $E_0$.

First, we comment on the absence of Kerr and Faraday rotations in our geometry. 
This is related to the fact that the off-diagonal $\epsilon_{yz}$  in eq.~(\ref{Eyz_txt}), 
which could have caused rotations in the $(y,z)$ plane, vanishes. 
Although time reversal symmetry is broken, there is no $yz$ Hall response. 
In a very different set up, where an external $\textbf{E}\!\cdot\!\textbf{B}$ term creates a charge imbalance between nodes,
one can get non-zero $\epsilon_{yz}$. This leads to a Faraday rotation~\cite{hosur:prb15}, which is independent of node separation 
and vanishes in the limit of equal Fermi energies at the two nodes.  

Next, we remark that the {\it longitudinal} component of $\textbf{E}$ obtained above is, in fact, just the
consequence of broken time reversal (and exists even in magnets~\cite{Visnovsky}); it is unrelated to topological properties of the TWS.
The latter can be probed in magnetic linear dichroism (MLD) measurements on surfaces with arcs.  
In cases (i) and (ii) discussed above, the waves propagate in the TWS with different refractive indices
$n_{\|}$ and $n_{\bot}$.
MLD measures the difference between absorption of light polarized linearly in different directions~\cite{Visnovsky}.
Thus the MLD signal yields
 \bea n_{\bot}''-n_{\|}'' \simeq\left(\frac{2\alpha}{n}\right)^3\left(\frac{b\lambda}{2\pi^2}\right)^2\frac{v_F}{c}.\eea 

\section*{Discussion}
We have presented in this paper detailed predictions for the Kerr and Faraday rotations and linear dichroism
in a Weyl semimetal. Our results give insights into various aspects of the TWS state
including its topological characteristics: Kerr and Faraday rotations from surfaces with no arcs and linear dichroism signal from surfaces supporting arcs. Perhaps the simplest candidate for testing our predictions 
is a topological insulator-trivial insulator multilayer geometry~\cite{burkov:prl11}, which has only two Weyl nodes.
For the TWS phase in the complex materials one must take into account multiple pairs of Weyl nodes. 
This can be done by a linear superposition of the results for individual pairs. 
Candidate materials for exploring the TWS phase include pyrochlore iridates A$_2$Ir$_2$O$_7$
in bulk~\cite{vishwanath:prb11,tokura:prl12} and in [111] thin films~\cite{chu:1309.4750,nagaosa:prl14},
and HgCr$_2$Se$_4$ spinels~\cite{xu:prl11,spinel:hall}. 

We should also note other systems whose ground state is not a TWS,
but where Weyl nodes can appear once time reversal is broken by an external magnetic field.
In the recently discovered Dirac semimetals Cd$_3$As$_2$\cite{Yi:Sciencereport14,Neupane:ncomm14,Jeon:nmat14} 
and Na$_3$Bi\cite{Li:science14,xu:science15,cava:Na3Bi}, each Dirac node is expected to split into two Weyl nodes
with a separation that grows linearly with ${B}$. 
On the other hand, in systems with a quadratic band touching~\cite{moon:prl13}, 
the application of a magnetic field leads to 
Weyl nodes with a separation proportional to $\sqrt{B}$.
Our results on Kerr, Faraday and linear dichroism signals
can also be generalized to these systems.

\section*{Methods}
We present here detailed derivations of many of the results reported in main text. 
We begin with the Kubo approach to the longitudinal and transverse conductivities of a TWS. 
We then turn to transmission and reflection matrices connecting scattered waves to incident waves. In the next two subsections, we discuss the modified Maxwell equations for a Weyl semimetal (TWS) and  reflection from a semi-infinite system. Finally, we give the details of the calculations for the case when light propagates perpendicular to node separation.

We use following notation in the text and Appendices. 
For all quantities $X= X' + iX''$ with $X' = {\rm Re}\, X$ and $X'' = {\rm Im}\, X$.
Subscripts $1,2$ etc.~refer to medium $1,2$ and {\it not to the real/imaginary parts}.
The permittivity $\tilde{\epsilon}$ and the permeability $\tilde{\mu}$ are defined via
the constitutive relations $\textbf{D}=\tilde{\epsilon}\textbf{E}$ and $\textbf{B}=\tilde{\mu}\textbf{H}$. 
Their ``relative'' counterparts, which are dimensionless, are defined by
$\epsilon=\tilde{\epsilon}/\epsilon_0$ and $\mu=\tilde{\mu}/\mu_0$, 
where $\epsilon_0(\mu_0)$ is permittivity (permeability) of vacuum. 
$\epsilon$ is the dielectric constant which then defines the refractive index as $n=\sqrt{\epsilon}$, 
and $\mu$ is connected to magnetic susceptibility $\chi_{m}$ as $\mu=1+\chi_{m}$. 
We assume that the susceptibility of the TWS is negligible.

\subsection*{Optical conductivity tensor of TWS\label{Opt_linear}}

In this subsection we calculate the longitudinal $\sigma_{xx}$ and transverse $\sigma_{xy}$ conductivities of a TWS. We consider the 
simplest case with only two nodes located at $\pm\textbf{b}$, where we choose $\textbf{b}=b\hat{z}$. 
Near the nodes, the linearized Hamiltonian is given by
\bea \label{Hw}
H(\textbf{k})=\pm \hbar v_F\vec{\sigma} \cdot (\bf{k}\pm\bf{b}).
\eea
The symbols used are defined in the main text. 
We obtain the conductivity from the Kubo formula 
\bea \label{kubo_para}
\sigma_{\alpha\beta}(\omega)=\frac{i}{\omega}\lim_{\textbf{q}\rightarrow 0}\Pi_{\alpha\beta}(\textbf{q},\omega).
\eea 
Note that there is no diamagnetic term for a strictly linear dispersion, a point which we will come back to below. 
The polarization function $\Pi_{\alpha\beta}(\textbf{q},\omega)$ is given by the current-current correlation function
\bea
\Pi_{\alpha\beta}(\textbf{q},i\omega_n)=\frac{-1}{\mathcal{V}}\int_{0}^{\beta}d\tau e^{i\omega_n\tau}
\langle T_{\tau} \hat{J}_{\alpha}(\textbf{q},\tau)\hat{J}_{\beta}(-\textbf{q},0) \rangle
\eea 
where $\mathcal{V}$ is the volume of the system and the current density operator 
\bea \label{J}
\hat{\bf{J}}=-\frac{\delta H}{\delta \bf{A}}=\pm ev_F \vec{\sigma}.
\eea 

The real frequency behavior is obtained by analytic continuation $i\omega_n\rightarrow \omega+i0^{+}$. 
For each node we obtain
\bea \nonumber 
\Pi_{\alpha,\beta}(\omega)=\frac{e^2v_F^2}{\mathcal{V}}\sum_{\textbf{k},i,i'}\frac{f(\varepsilon_{i'}(\textbf{k}))-f(\varepsilon_{i}(\textbf{k}))}{\hbar{\omega}+\varepsilon_{i'}(\textbf{k})-\varepsilon_{i}(\textbf{k})+i0^{+}}
\langle \textbf{k}i|\sigma_{\alpha}|\textbf{k}i'\rangle \langle \textbf{k}i'|\sigma_{\beta}|\textbf{k}i\rangle,
\eea
where $f(x)=1/(e^{\beta x}+1)$ is the Fermi function. 
The quasiparticle energies and eigenstates are obtained from $H(\textbf{k})|\textbf{k}i\rangle=\varepsilon_{i}(\textbf{k})|\textbf{k}i\rangle$ with $i=1, 2$ labeling two eigenstates at each wavevector.  

We evaluate the longitudinal and transverse polarizations $\Pi_{\alpha\beta}(\omega)=\Pi_{\alpha\beta}'(\omega)+i\Pi_{\alpha\beta}''(\omega)$ when the Fermi energy lies with nodes. 
Therefore, at the low frequency limit $\omega<\omega_c$ the total longitudinal conductivity from both nodes is given by:  
\bea \label{real_xx} &&\sigma_{xx}'(\omega)=\frac{e^2}{6\pi h}\frac{\omega}{v_F},\\ 
\label{im_xx} &&\sigma_{xx}^{''}(\omega)=-\frac{e^2}{3\pi^2\hbar v_{F}}\left\{\frac{\omega_c^2}{\omega}+\frac{\omega}{4}\ln\left|\frac{\omega^2-4\omega_c^2}{\omega^2}\right|\right\}.\eea 

Note that the spurious $1/\omega$ term in $\sigma_{xx}^{''}$ is an artifact of effective linear Hamiltonian with no diamagnetic response. 
We have shown that this problem is cured when we consider a lattice Hamiltonian~\cite{yingran:prb11,Hughes:arxiv1405.7377} 
 \bea \label{H_lattice} \nonumber 
 H(\textbf{k})=2t\sin(k_{x}a)\sigma_{x}+ 2t\sin(k_{y}a)\sigma_{y}+2t[2+\cos(ba)]\sigma_z -2t\left[\cos(k_{x}a)+\cos(k_{y}a)+\cos(k_{z}a)\right]\sigma_{z}, \eea 
 where $a$ is the lattice spacing.
For the lattice model we found that the diverging $1/\omega$ terms in paramagnetic and diamagnetic responses precisely cancel each other.
Therefore, in the properly regularized model, we are left only with log term in the equation above for $\sigma_{xx}^{''}(\omega)$ with
a cutoff that is effectively $\omega_c \sim t \sim \hbar v_F/a$. This lead to a log term in the real part of dielectric constant $\epsilon'(\omega)$. 

In the limit of low frequencies, we find the transverse conductivity is given by:
\bea \label{real_xy} \sigma_{xy}'(\omega)=\frac{e^2}{\pi h}b+\frac{e^2}{6v_F^2\pi h}\frac{b}{k_c^2-b^2}\omega^2,\eea where 
we denote here by $k_c$, the cut-off along $k_z$ axis.   

\subsection*{Transmission and reflection coefficients for ultra-thin films\label{TR_Coeff}}  
In this subsection we derive the reflection and transmission matrices in the ultra-thin film limit
where the TWS film thickness $d \ll \lambda$, the wavelength of light, even though $d \gg a$, the lattice constant.
Thus the two media $i$ ($z <0$) and $j$ ($z >0$) are separated by a TWS which gives rise to the boundary conditions at $z =0$.

Taking the incident wave to propagate along the $z$ axis in medium $i$ as $\textbf{E}^{in}=(E_x^{i}\hat{x}+E_y^{i}\hat{y})e^{ik_iz}$, the expressions for EM waves in 
the two media are as 
\bea \nonumber 
&\mathrm{Medium}~i\hspace{-0.1cm}:~\textbf{E}_i=e^{ik_iz}\left(
 \begin{array}{cccc}
  E_x^i\\ 
 E_y^i\\  
  \end{array}\right)+e^{-ik_iz}\left(
  \begin{array}{cccc}
  E_x^{ri}\\ 
 E_y^{ri}\\  
  \end{array}\right),\\ 
& \mathrm{Medium}~j\hspace{-0.1cm}:~\textbf{E}_j=e^{ik_jz}\left(
  \begin{array}{cccc}
  E_x^{tj}\\ 
 E_y^{tj}\\  
  \end{array}\right),
\eea
where wavevector is $k_i=\omega n_{i}/c$ using notation $n_i=\sqrt{\epsilon_i}$, and superscripts $r$ and $t$ stand for reflection and transmission components.  
We assume medium $i$ is vacuum $n_i=1$ and medium $j$ (substrate) has refractive index $n_j=n$. In the main text we only quote the results for a 
free-standing film, i.e., for $n_j = 1$.

The EM waves should match on the boundary and are given by $\textbf{n}\times (\textbf{E}_i-\textbf{E}_j)=0$ and $\textbf{n}\times (\textbf{H}_i-\textbf{H}_j)=\textbf{J}_{s}$, where $\bf{n}=-\hat{z}$ is normal to interface.
The surface current density $\bf{J}_s=\sigma^{s}\textbf{E}$, where the surface conductivity tensor
$\sigma^{s}_{\alpha\beta}=d \sigma_{\alpha\beta}$ with the bulk $\sigma_{\alpha\beta}$ for the TWS given by the results in
Appendix~\ref{Opt_linear}.

Solving the boundary value problem, we get the following expressions for transmitted wave 
\bea \label{Ext}
E_{x}^{t}=\frac{2(1+n+\frac{1}{c\epsilon_0}\sigma^s_{xx})}{(1+n+\frac{1}{c\epsilon_0}\sigma^s_{xx})^2+(\frac{1}{c\epsilon_0}\sigma^s_{xx})^2}E_{x}^{i},\\ \label{Eyt}
E_{y}^{t}=\frac{ \frac{2}{c\epsilon_0}\sigma^s_{xy}}{(1+n+\frac{1}{c\epsilon_0}\sigma^s_{xx})^2+(\frac{1}{c\epsilon_0}\sigma^s_{xx})^2}E_{x}^{i}.
\eea 
We use these results to calculate Faraday rotation in Eq.~(2) in main text. 
Similarly, for reflection components we have
\bea \label{Exr}
E_{x}^{r}&=&\frac{1-\left(n+\frac{1}{c\epsilon_0}\sigma^s_{xx}\right)^2-\left(\frac{1}{c\epsilon_0}\sigma^s_{xy}\right)^2}{(1+n+\frac{1}{c\epsilon_0}\sigma^s_{xx})^2+(\frac{1}{c\epsilon_0}\sigma^s_{xx})^2}E_{x}^{i},\\ \label{Eyr}
E_{y}^{r}&=&\frac{\frac{2}{c\epsilon_0}\sigma^s_{xy}}{(1+n+\frac{1}{c\epsilon_0}\sigma^s_{xx})^2+(\frac{1}{c\epsilon_0}\sigma^s_{xx})^2}E_{x}^{i}.
\eea 
These relations are used to calculate the Kerr rotation in Eq.~(3).
        
\subsection*{Modification of Maxwell equations in TWS \label{Mod_Maxwel}}
The low energy electromagnetic response of Weyl semimetal is described by an spatially varying axion term. Including conventional Maxwell term, the full action 
$S=S_0+S_\theta$ of system is~\cite{Qi:prb08} 
\bea 
\nonumber &&S_0=\int d^3xdt\left\{-\frac{1}{4\mu_0}F_{\mu\nu}F^{\mu\nu}+\frac{1}{2}F_{\mu\nu}\mathcal{P}^{\mu\nu}-J^{\mu}A_{\mu} \right\},\\ \label{S_axion}
&&S_{\theta}=\frac{\alpha}{8\pi\mu_0}\int d^3xdt\left\{\theta(\textbf{r},t)\epsilon^{\mu\nu\alpha\beta}F_{\mu\nu}F_{\alpha\beta}\right\}.
\eea 
Here the tensor $\mathcal{P}^{\mu\nu}$ stand for electric polarization and magnetization as $\mathcal{P}^{0i}=cP^{i}$ and $\mathcal{P}^{ij}=-\epsilon^{ijk}M_{k}$.
The axion field $\theta$ varies with space and time as $\theta(\textbf{r},t)=2\textbf{b}\cdot \textbf{r}-2b_{0}t$, where 
$\textbf{b}$($b_0$) denotes separation of nodes in momentum (energy) space~\cite{zyuzin:prb12}. 
We set  $b_0$ to be zero, since in the problem of our interest both the Weyl nodes are at the same chemical potential. 
 
Varying action with respect to $A_{\mu}$, the equation of motions are obtained as follows. \bea -\frac{1}{\mu_0}\partial_{\nu}F^{\mu\nu}+\partial_{\nu}\mathcal{P}^{\mu\nu}+\frac{\alpha}{2\pi\mu_0}\epsilon^{\mu\nu\alpha\beta}\partial_{\nu}(\theta F_{\alpha\beta})=J^{\mu}.\eea
These equation of motions yield the modified Maxwell equations in the presence of axion field in the effective action~\cite{wilczek:prl87}.
In a more conventional form, they are as 
\bea \label{E}
&&\nabla\times\textbf{E}+\frac{\partial \textbf{B}}{\partial t}=0,\\ \label{B}&&\nabla\cdot \textbf{B}=0, \\ \label{D}&&\nabla\cdot \textbf{D}=\rho+\frac{2\alpha}{\pi}\sqrt{\frac{\epsilon_0}{\mu_0}}\textbf{b}\cdot \textbf{B}, \\ \label{H_modified}
\label{H}&&\nabla\times\textbf{H}=\frac{\partial \textbf{D}}{\partial t}+\textbf{J}-\frac{2\alpha}{\pi}\sqrt{\frac{\epsilon_0}{\mu_0}} \textbf{b}\times \textbf{E},
\eea   

In the absence of topological terms proportional to $\textbf{b}$, we assume that the constitutive relations are isotropic 
with $\textbf{D}=\tilde{\epsilon}\textbf{E}$ and $\textbf{B}=\tilde{\mu}\textbf{H}$. Here 
$\epsilon=\tilde{\epsilon}/\epsilon_0=\epsilon_b+i\sigma_{l}/\epsilon_0\omega$ and $\mu=\tilde{\mu}/{\mu_0}=1+\chi_m$, where $\epsilon_b$ denotes the bound charge contribution and $\sigma_l$ is longitudinal conductivity. With the topological term, the dielectric constant acquires
off-diagonal elements, and is given by
\bea\label{e_tensor} \overleftrightarrow{\epsilon}=\epsilon\textbf{1}-\frac{2\alpha}{\pi}\frac{c}{\omega}b\tau^{y}, \eea 
where $\tau^{y}$ is Pauli matrix that comes from writing $\textbf{b}\times \textbf{E}$ in terms of its components when $\textbf{b}=b\hat{z}$.   
  
\subsection*{Reflection from a semi-infinite system \label{semi_infinite}} 
We consider an interface between two media, say vacuum and TWS, and match the incoming, the reflected and transmitted waves as follows
\bea 
&&\textbf{E}_{r}=\left(\textbf{1}+\mathcal{N}\right)^{-1}\left(\textbf{1}-\mathcal{N}\right)\textbf{E}_{in},\\ 
&&\textbf{E}_{t}=2\left(\textbf{1}+\mathcal{N}\right)^{-1}\textbf{E}_{in},
\eea
where we have defined \bea \label{N} \mathcal{N}=\sqrt{\frac{\mu_1\epsilon_2}{\epsilon_1\mu_2}}\left(\textbf{1}-\frac{2\alpha}{\pi}\frac{c}{\omega}\frac{1}{\epsilon_2}~b\tau^{y}\right)^{1/2}.\eea 
For vacuum (medium 1) $\epsilon_{1}=\mu_{1}=1$ and for Weyl semimetal (medium 2) we assume $\epsilon_{2}=\epsilon$ and $\mu_2=1$ in following.

Assuming incoming wave is polarized along $x$ axis as $\textbf{E}_{in}=E_0\hat{x}$, 
we obtain 
\bea
 \left(\begin{array}{cccc}
  E^r_x\\ 
 E^r_y\\  
  \end{array}\right)=R \left(\begin{array}{cccc}
 E_0\\ 
 0\\  
  \end{array}\right),
  \left(\begin{array}{cccc}
  E^t_x\\ 
 E^t_y\\  
  \end{array}\right)=T \left(\begin{array}{cccc}
 E_0\\ 
 0\\  
 \end{array}\right).
\eea  
The reflection and transmission matrices are given by
\bea \nonumber R=A\left(\begin{array}{cccc}
  1-n_{+}n_{-}&i(n_{+}-n_{-})\\ 
 i(n_{-}-n_{+})&1-n_{+}n_{-}\\  
 \end{array}\right),~~
 T=A\left(\begin{array}{cccc}
  2+n_{+}+n_{-}&i(n_{+}-n_{-})\\ 
 i(n_{-}-n_{+})&2+n_{+}+n_{-}\\  
 \end{array}\right),\eea 
 where $A=1/(1+n_{+})(1+n_{-})$ and 
 \bea \label{npm_app} n_{\pm}=\sqrt{\epsilon_2 \pm \frac{2\alpha}{\pi}\frac{c}{\omega}b },
 \eea 
 which is Eq.~(7) in the main text. The reflection matrix $R$ is used to calculate the Kerr rotation as follows. Defining $(n_{-}-n_{+})/(1-n_{-}n_{+})=\eta e^{i\phi}$ The electric field on reflection becomes  
 \bea \textbf{E}=E_{0}(\hat{x}+i\eta e^{i\phi} \hat{y})e^{i(kz-\omega t)}.\eea The electric field is elliptically polarized and major axis is rotated by $\theta_{K}$, where 
 \bea \tan 2\theta_{K}=\frac{2\eta \sin\phi}{\eta^2-1}. \eea  
  
\subsection*{Propagation of light perpendicular to node separation\label{arc_maxwell}}
We first analyze the propagation of light in WSM assuming the direction of propagation is perpendicular to node separation as shown in Fig.~1(c). Although the Figure shows the incident ${\bf E}$ is polarized along $y$ axis, to begin with we consider a more general case, but still keeping the $x$-axis as the direction of propagation. 
Our goal is to derive Eq.~(11)  in the main text.

Using modified Maxwell equations~(\ref{E}) and (\ref{H}), we obtain
\bea 
\nabla\times\nabla\times\textbf{E}=-\frac{\epsilon}{c^2}\frac{\partial^2\textbf{E}}{\partial t^2}+\frac{2\alpha}{\pi c}\textbf{b}\times\frac{\partial\textbf{E}}{\partial t}. 
\eea 
Taking the electric field to be as $\textbf{E}(\textbf{r},t)=\textbf{E}e^{i(kx-\omega t)}$ with wavevector $k=n\omega/c$ and using the 
identity $\nabla\times\nabla\times\textbf{E}=\nabla \nabla\cdot\textbf{E}-\nabla^2\textbf{E}$, we obtain Eq.~(11). 
 
Now we specialize to the case where ${\bf E}$ is polarized along $y$ axis (case (ii) in the main text), whose geometry is shown in Fig.~1(c).
We must impose the boundary conditions on the tangential ${\bf E}$-fields,
$E_{y}^{in}+E_{y}^{r}=E_{y}^{t}$, and ${\bf H}$-fields,
$E_{y}^{in}-E_{y}^{r}-nE_{y}^{t}=\sqrt{\mu_0/\epsilon_0}J^{s}_y$.

We must carefully consider what contributes to the surface current density $J^{s}_y$.
The bulk conductivity $\sigma_{yy}$ is not responsible for 
surface currents. As shown in the main text, the longitudinal current density in a WSM is not localized at the surface.
In contrast to good conductors, a WSM has a very long penetration depth $\delta\sim\epsilon'\lambda/\pi\epsilon''\gg\lambda$.

The only contribution to the surface current density is then $J^{s}_y=\sigma^{s}_{yx}E_x$.
Here $E_x(x=0)$ arises from the longitudinal field in the WSM already described in the main text
and the off-diagonal $\sigma^{s}_{yx}$ arises from the unusual surface states. 
To compute $\sigma^{s}_{yx}$, we look at the contribution of an individual surface state, labeled by the wavevector $k_z$,
and sum up all the contributions. The surface state labelled by $k_z$ is localized near the $x=0$ plane 
with a localization length~\cite{burkov:prl11,okugawa:prb14} given by $\xi(k_{z})={2b}/{(b^2-k_z^2)}$.
As one approaches the tips of the Fermi arc at $k_z = \pm b$, the localization length diverges, and the surface state connects as it were with the bulk states. 
Thus to compute the surface conductivity we must cut off the $k_z$'s on the scale of the wavelength of light.  
Using the fact that each individual ``layer" in k-space labeled by $k_z$ yields a 2D quantized Hall conductivity 
$\sigma^{2D}(k_z)=e^2/h$, we obtain the surface conductivity 
\bea 
\label{sigmas_appE} \sigma^{s}=\int_{\xi\le\lambda} \frac{dk_z}{2\pi}\sigma^{2D}(k_z)\xi(k_z)\simeq\frac{e^2}{\pi h}\ln(2b\lambda).
\eea 

Uisng the surface conductivity in eq.~(\ref{sigmas_appE}) in the boundary conditions, 
we can calculate the reflection $r=E^r_{y}/E^{in}_{y}$ and transmission $t=E^t_{y}/E^{in}_{y}$ coefficients. 
They are are found to be $r=(1-\bar{n})/(1+\bar{n})$ and $t=2/(1+\bar{n})$, 
where the generalized refractive index is given by $\bar{n}=n+2{\alpha^2}b\lambda\ln(2b\lambda)/{\pi^3n}$. 
Therefore, the $t$ and $r$ coefficients are functions of the node separation $b$.


\section*{Acknowledgements }
We acknowledge the support of the CEM, an NSF MRSEC, under grant DMR-1420451. 
We thank Y-M. Lu and R. Valdes Aguilar for useful discussions.
M.K. would also like to thank P. Hosur and J. H. Wilson for discussions. 

\section*{Author contributions statement}
M.K., M.R. and N.T. contributed to the theoretical research described in this paper and the writing of the manuscript. 

\section*{Additional information}
Competing financial interests: The authors declare no competing financial interests.

\end{document}